# Observation of charge density wave order in 1D mirror twin boundaries of single-layer MoSe$_2$

Sara Barja[1,2]★*, Sebastian Wickenburg[1,2]★, Zhen-Fei Liu[1,2], Yi Zhang[3,4], Hyejin Ryu[3], Miguel M. Ugeda[5,6,7], Zahid Hussain[3], Z.-X. Shen[8], Sung-Kwan Mo[3], Ed Wong[1,2], Miquel B. Salmeron[2,5,9], Feng Wang[2,5,10], Michael F. Crommie[2,5,10], D. Frank Ogletree[1,2]*, Jeffrey B. Neaton[1,2,5,10], Alexander Weber-Bargioni[1,2]*

[1] Molecular Foundry, Lawrence Berkeley National Laboratory, California 94720, USA. [2] Materials Sciences Division, Lawrence Berkeley National Laboratory, California 94720, USA. [3]Advanced Light Source, Lawrence Berkeley National Laboratory, Berkeley, CA 94720, USA. [4]National Laboratory of Solid State Microstructures, School of Physics, Collaborative Innovation Center of Advanced Microstructures, Nanjing University, Nanjing 210093 P.R.China.[5]Department of Physics, University of California at Berkeley, Berkeley, California 94720, USA.[6]CIC nanoGUNE, Donostia-San Sebastian 20018, Spain.[7]Ikerbasque, Basque Foundation for Science, Bilbao 48011, Spain. [8]Stanford Institute of Materials and Energy Sciences, SLAC National Accelerator Laboratory, Menlo Park, California 94025, USA. [9]Department of Materials Science and Engineering, University of California Berkeley, 210 Hearst Mining Building, Berkeley, California 94720, USA. [10]Kavli Energy NanoSciences Institute at the University of California Berkeley and the Lawrence Berkeley National Laboratory, Berkeley, California 94720, USA

★These authors contributed equally to this work

* e-mail: sbarja@lbl.gov ; dfogletree@lbl.gov ; afweber-bargioni@lbl.gov

**Properties of two-dimensional transition metal dichalcogenides are highly sensitive to the presence of defects in the crystal structure. A detailed understanding of defect structure may lead to control of material properties through "defect engineering". Here we provide direct evidence for the existence of isolated, one-dimensional charge density waves at mirror twin boundaries in single-layer $MoSe_2$. Our low-temperature scanning tunneling microscopy/spectroscopy measurements reveal a substantial bandgap of 60 - 140 meV opening at the Fermi level in the otherwise one dimensional metallic structure. We find an energy-dependent periodic modulation in the density of states along the mirror twin boundary, with a wavelength of approximately three lattice constants. The modulations in the density of states above and below the Fermi level are spatially out of phase, consistent with charge density wave order. In addition to the electronic characterization, we determine the atomic structure and bonding configuration of the one-dimensional mirror twin boundary by means of high-resolution non-contact atomic force microscopy. Density functional theory calculations reproduce both the gap opening and the modulations of the density of states.**

Intrinsic defects play an important role in determining and modifying the properties of two-dimensional (2D) materials, and have been studied extensively in graphene[1–5]. Defects in atomically thin 2D semiconductors, on the other hand, have been explored to a lesser extent, but are expected to substantially modify 2D material properties. 2D transition metal dichalcogenide (2D-TMD) semiconductors are particularly interesting since they exhibit direct bandgaps in the visible range[6–8], high charge-carrier mobility[9,10], extraordinarily enhanced light-matter interactions[11–14] and potential applications in novel optoelectronic devices[15,16]. Individual atomic scale defects in 2D-TMDs have been predicted to modify charge transport[17] or introduce ferromagnetism[18], while one dimensional (1D) defects such as grain boundaries and edges may alter electronic[19] and optical properties[19,20], and introduce magnetic[21] and catalytic[22,23] functionality. Controlled incorporation of defects, or "defect engineering", in 2D semiconductors may allow tailoring of material properties for diverse applications.

Here we report the direct observation of one-dimensional charge density waves (CDWs) at mirror twin boundaries in monolayer $MoSe_2$, a 2D-TMD semiconductor. A 1D CDW is a macroscopic quantum state, where atoms in a 1D metallic system relax and break translational symmetry to reduce electronic energy by opening a small bandgap at the Fermi energy and modulating the charge density at the periodicity of the lattice distortion[24,25]. While CDW order has been observed in 2D-TMD metals such as $NbSe_2$ and $TiSe_2$ at low temperature[26,27], CDWs have not previously been associated with 2D-TMD semiconductors.

Most studies of one-dimensional charge density waves have been performed on ensembles of CDWs in conducting polymers, quasi one-dimensional metals or self-assembled atomic chains adsorbed on semiconducting surfaces, where inter-CDW coupling can significantly impact CDW properties[28–32]. The CDWs observed here are electronically isolated from one another and have truly one-dimensional character, forming an atomically precise model system to explore intrinsic CDW phenomena. We observed CDWs along 1D networks of mirror twin boundaries (MTBs) in monolayer crystals of $MoSe_2$ grown by molecular beam epitaxy. This particular type of MTB has been observed using transmission electron microscopy and predicted to be metallic in $MoSe_2$ and $MoS_2$[19,33–37]. Various morphologies have been proposed for mirror twin boundaries in 2D semiconductors[19,34,35,38], and scanning tunneling microscopy experiments have identified 1D metallic wires along boundaries in single-layer of $MoSe_2$[36,39].

We used low temperature (4.5K) scanning tunneling microscopy/spectroscopy (STM/STS) to detect the bandgap of a CDW along mirror twin boundaries in single-layer $MoSe_2$ and to spatially map the modulation of the charge density. We determined the atomic structure of the MTB with parallel high-resolution non-contact atomic force microscopy imaging. Based on this MTB structure model, we investigated the electronic structure by first principles density functional theory (DFT) calculations.

Fig. 1a shows a high-resolution STM image of single-layer $MoSe_2$ grown by molecular beam epitaxy on bilayer graphene (BLG) on SiC(0001). Sharply defined pairs of identical parallel lines decorate the $MoSe_2$ monolayer. These bright lines have previously been described as inversion domain boundaries[36]. The apparent height of the features strongly depends on the bias applied between tip and sample, implying that the STM contrast originates mainly from the locally modified electronic structure rather than a protruding topographic feature. High-resolution images of the boundary show that the contrast is modulated along the defect line (Fig. 1b). The modulation period is ~1 nm, approximately three unit cells of $MoSe_2$. This modulation has previously been attributed to a superlattice potential induced by the nearly commensurate 3 x 3 moiré pattern that forms when a monolayer of $MoSe_2$ is atomically aligned with the graphite substrate[36]. We are able to rule out this explanation because the periodicity along the boundary remained constant (~1 nm) for different TMD-graphene registries and associated moiré pattern periodicities (Supplementary).

We mapped the local density of states (LDOS) using STM dI/dV spectroscopy to understand the MTB contrast. The inset of Fig. 1c shows a dI/dV spectrum measured on pristine monolayer $MoSe_2$ away from the MTB, showing an electronic bandgap of 2.18 eV in good agreement with previous measurements[14]. In contrast, dI/dV spectra acquired on the MTB show numerous features within the semiconducting bandgap (Fig. 1c). A high-resolution view of this spectrum around the Fermi level (Fig. 2a) reveals a bandgap of 73 mV with sharp peaks at the band edges. The bandgap for each MTB is constant along its entire length, but different MTBs show experimental gaps with an average of 95 mV ± 10 mV. The dI/dV spectrum also shows a set of satellite peaks adjacent to the band edge peaks, offset by 14.2 ± 0.8 meV for both occupied and unoccupied states.

To investigate the nature of the MTB bandgap, we studied the spatial distribution of the electronic band edge states ($\psi_-$ and $\psi_+$ in Fig. 2a) by measuring constant-height dI/dV conductance maps. Fig. 2b shows representative dI/dV maps of the edge states below ($\psi_-$) and above ($\psi_+$) the gap characterized by a strong modulation with a periodicity of ~1nm. Fig. 2c also shows that the occupied (blue) and unoccupied (red) state modulations are spatially out of phase with nearly constant amplitude along the MTB. We can exclude drift-related artifacts between the two conductance maps because they were obtained simultaneously by recording the forward and backward scan lines at the two different biases.

While high-resolution STM images reveal the atomic structure of the MoSe$_2$ layer away from the domain boundary (Fig. 3a), the significant electronic contribution to the STM contrast near the line defect (Fig. 3b) makes it impossible to determine the MTB atomic structure. Therefore, we performed non-contact atomic force microscopy (nc-AFM) using CO-functionalized tips[40] for enhanced spatial resolution. Since nc-AFM frequency shifts are not affected by the electronic structure near the Fermi energy, we were able to obtain atomically resolved images of the MTB. Fig. 3c shows the nc-AFM frequency shift image taken at the same location and with the same tip as the STM image in Fig. 3a, identifying unambiguously the precise atomic morphology of the domain boundary.

To distinguish between Mo and Se atoms we first identify the well known atomic structure away from the MTB, which reveals a hexagonal lattice of bright features. At the small tip-sample separation used here, bright areas (higher frequency shift) in nc-AFM are generally due to short-range repulsive Pauli forces, while dark areas (lower frequency shift) are due to long range attractive van der Waals and electrostatic forces. Following this intuitive picture, we attribute the bright features (yellow) to the higher-lying Se atoms, which are close enough to the tip to generate repulsive forces, and the dark features (blue) to the lower lying Mo atoms, whose larger distance from the tip leads to only attractive forces. The MTB consists of an atomic line of Se atoms surrounded by a darker region of stronger attractive forces. The stronger dark contrast is due to a higher density of Mo atoms, where each Se atom is bound to four Mo atoms instead of three. Simulations of the image contrast using previously established methods[41] (overlaid in the upper part of the nc-AFM image in Fig. 3c) supports this structural model. The Se atoms form a continuous hexagonal lattice across the boundary, while the Mo atom positions are reflected across the single atom wide line defect (Fig. 3d), forming a mirror twin boundary[33–36,38]. All of the MTB structures in our samples are identical.

Due to the finite length of the MTBs, the observed modulations in their intensity have been discussed in terms of quantum well states[36]. This cannot explain the LDOS features we observed (Fig. 2b), since MTBs as long as 30 nm show uniform LDOS modulation, whereas the modulation amplitude should decay away from the ends of the MTB in the case of quantum confinement. We believe that our results are best explained by the presence of a charge density wave since a) we observe a bandgap in an otherwise metallic electronic structure, b) the edge states' periodicity is nearly three times the lattice constant, and c) occupied and unoccupied states are spatially out of phase. In order to test this hypothesis we investigated the electronic structure of the MTB by means of DFT calculations.

Starting with the measured atomic structure of the MTB, we computed the electronic structure using DFT employing the Perdew-Burke-Ernzerhof (PBE) functional[42]. Using one unit cell along the MTB (grey area in Fig. 4a), consistent with those used in prior work in both $MoSe_2$[33] and $MoS_2$[34,38], the MTB gives rise to a metallic band that crosses the Fermi energy at about 1/3 of the way along the Γ-X direction of the Brillouin zone (Supplementary). This band is absent in pristine monolayer $MoSe_2$. We studied the effect of the spin-orbit coupling on the electronic structure and found neither substantial modification of the band structure, although the degeneracy of states around the midpoint of the Brillouin zone is lifted (Supplementary).

The one-dimensional nature of the MTB, combined with a dispersive band crossing the Fermi level at around 1/3 of the Brillouin zone, suggests the possibility of a CDW. We explored the CDW by tripling the size of our system, using a larger supercell with three unit cells along the MTB, which is close to the experimentally measured periodicity. We did not observe the spontaneous formation of a CDW in the calculation, which may be related to the filling of the calculated band being close to, but less than 1/3 (Supplementary). We do not believe that electron-electron interactions produce the CDW ordering observed here, because of the significant dispersion of the metallic band and the screening from the nearby material and substrate. An alternative mechanism for CDW formation involves a Peierls instability[25], where the lattice is periodically distorted to reduce the total energy of the system by opening a gap at the Fermi level and leading to charge density modulation with a periodicity that reflects the lattice distortion. We therefore introduced lattice distortions to explore the effects of symmetry breaking on the electronic structure. Since any distortions examined theoretically have to be necessarily commensurate with the simulation cell we introduced distortions with a periodicity of three times the lattice constant along the MTB, consistent with the experimentally measured periodicity. Fig. 4a illustrates one such distortion, where the red arrows indicate the displacements of 0.05 Å along the line defect. As seen in Fig. 4b this is sufficient to open a bandgap of ~50 meV at the edge of the folded Brillouin zone (the magnitude of the gap is linear with the distortion amplitude). Since the Fermi wavevector is slightly below the edge of the folded Brillouin zone, the gap opens not at the Fermi energy, but slightly above (0.175 eV in this case). After aligning the bandgaps, we get excellent agreement between the calculated

DOS and the experimental dI/dV spectrum, including features as far away as 0.5 eV from $E_F$ (Fig. 4c). The spatial extent of the calculated LDOS also matches the experimental data as seen in Fig. 4d. Here the $\psi-$ ($\psi+$) states below (above) the gap are seen to exhibit the same shape, periodicity, and spatial phase shift as observed in our measurements. States at smaller wavevectors, away from the folded zone boundary, do not exhibit this modulation in our DFT calculations (Supplementary). Additionally, DFT-PBE calculations are well known to underestimate the bandgap. Thus, an actual distortion along the boundary for the same gap will exhibit a somewhat smaller amplitude than predicted here, which is below the resolution of our nc-AFM measurements. We have considered different types of distortions and the results are all qualitatively similar in opening a gap near the edge of the folded Brillouin zone. Crucially, distortions of two times the lattice periodicity in super cells containing two unit cells along the MTB do not open a gap in the density of states near the Fermi level (Supplementary). Taken together, our experimental and theoretical results show strong evidence for the formation of a one-dimensional CDW along the MTB.

In conclusion, our combined STS/nc-AFM measurements, supported by DFT calculations, allowed us to fully characterize the atomic and electronic structure of an isolated one-dimensional charge density wave. The $MoSe_2$ MTB provides an exceptional model to study the physics of symmetry breaking in one-dimensional correlated systems. It also presents new opportunities to study the charge transport in such systems, including Fröhlich conduction, as well as the effects of pinning due to point defects or adatoms, and confinement due to intersections between macroscopically interconnected CDWs within a one-dimensional network.

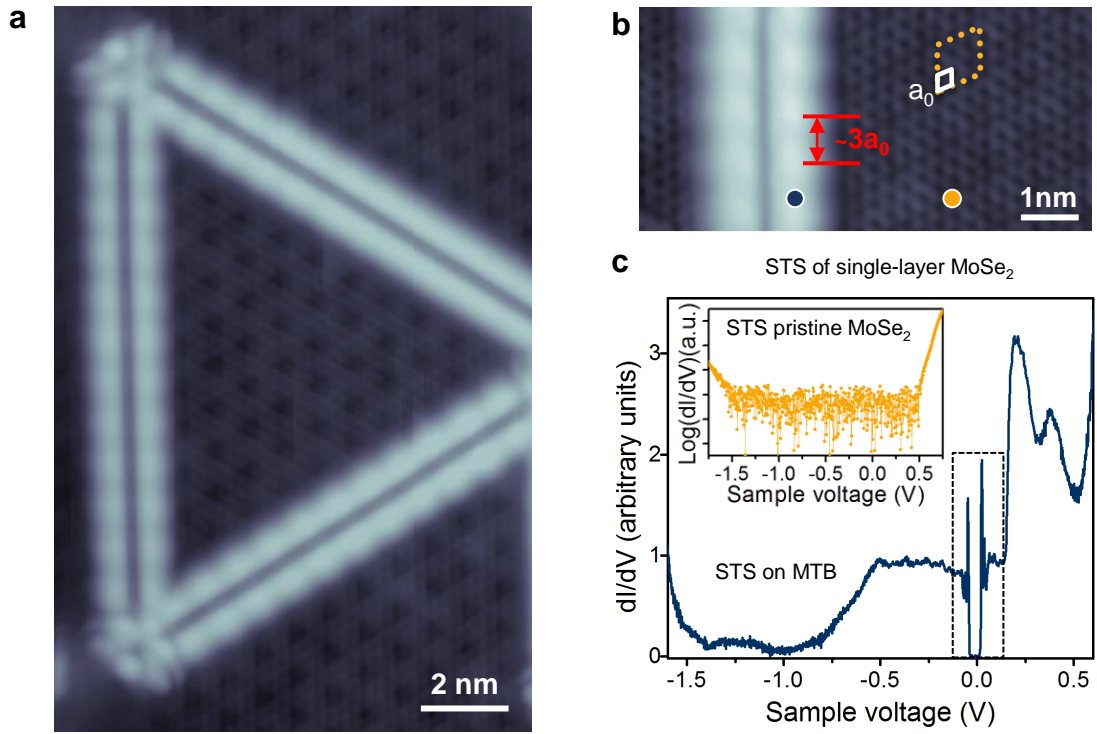

**Figure 1: Morphology and electronic structure of mirror twin boundaries in monolayer MoSe$_2$**. **a**, Large scale STM image of single-layer MoSe$_2$ on bilayer graphene (V$_s$= -1.5 V, I$_t$= 10 pA, T= 4.5 K). **b**, High-resolution image of a mirror twin boundary in single-layer MoSe$_2$ (V$_s$= -1.5 V, I$_t$= 10 pA, T= 4.5K) showing a 3x3 moiré pattern (dashed yellow line, with a single unit cell shown in white) and a modulation of the electronic states near the MTB with a wavelength of ~3a$_0$. **c**, Typical STM dI/dV spectra acquired on a mirror twin boundary (blue) and on the pristine monolayer MoSe$_2$/BLG (inset - yellow) show that the MTB has states throughout the semiconducting gap of monolayer MoSe$_2$. The dashed box indicates a band gap opening around the Fermi level.

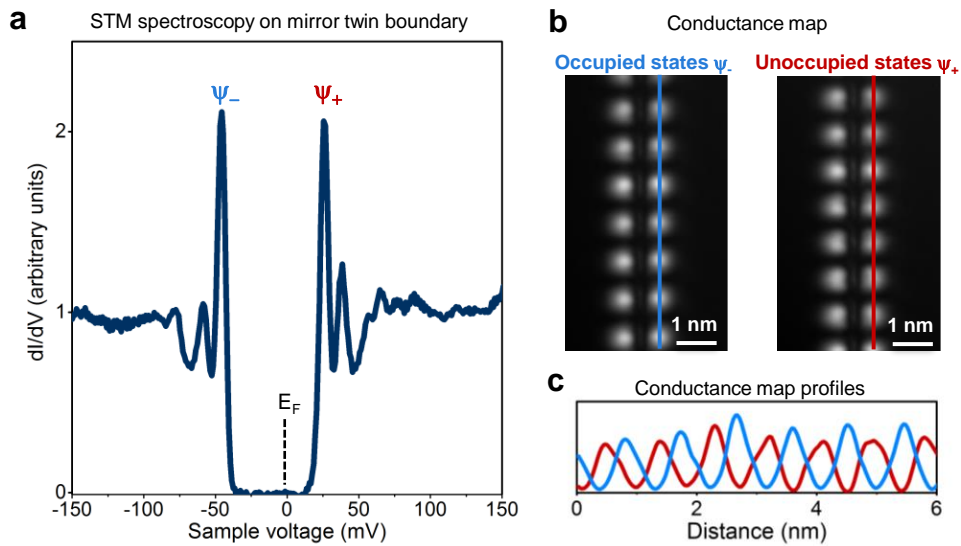

**Figure 2: Local density of states of mirror twin boundaries with an energy gap around the Fermi level. a**, High resolution STM dI/dV spectrum of a mirror twin boundary on monolayer MoSe$_2$/BLG showing an energy gap of 73 meV. 0 V sample voltage represents E$_F$ **b**, Representative dI/dV constant-height conductance maps recorded at voltages corresponding to edges states below ($\psi_-$) and above ($\psi_+$) the gap. Bright regions indicate larger DOS. **c**, Line profile of the dI/dV maps in b for occupied (blue) and empty (red) states, showing that the maxima for two states are spatially out of phase.

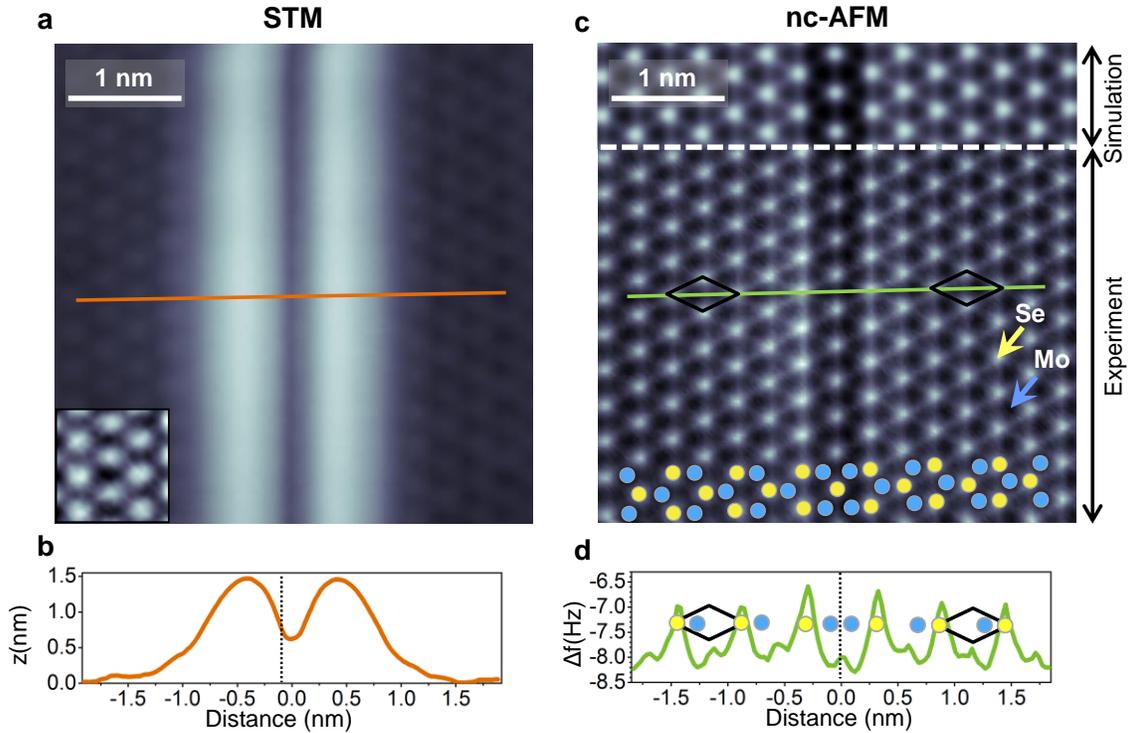

**Figure 3: Atomic resolution nc-AFM image of a mirror twin boundary reveals the precise atomic structure. a**, STM topographic image of a mirror twin boundary in monolayer MoSe$_2$/BLG ($V_s$ = -1.5 V, $I_t$ = 10 pA, T = 4.5 K). Inset: cut-off of the main STM image, where the color contrast has been adjusted to resolve the atomic lattice. **b**, Line profile of the STM image (orange line) showing the significant electronic contribution to the STM signal near the MTB. **c**, nc-AFM image measured in same area and with the same tip as **a**. Se (Mo) atom positions are marked by yellow (blue) circles at the bottom of the figure. The Se atoms (bright areas, yellow) form an uninterrupted hexagonal lattice, whereas the Mo atoms (dark areas, blue), switch their positions within the indicated unit cells of the Se (black diamonds) between the two sides of the MTB. This assignment is corroborated by simulation of the frequency shift image (upper part of the figure above white dashed line). **d**, Line profile of the nc-AFM image (green line) with the Se (yellow) and Mo (blue) atom positions overlaid, showing the mirror symmetry of the Mo lattice across the MTB.

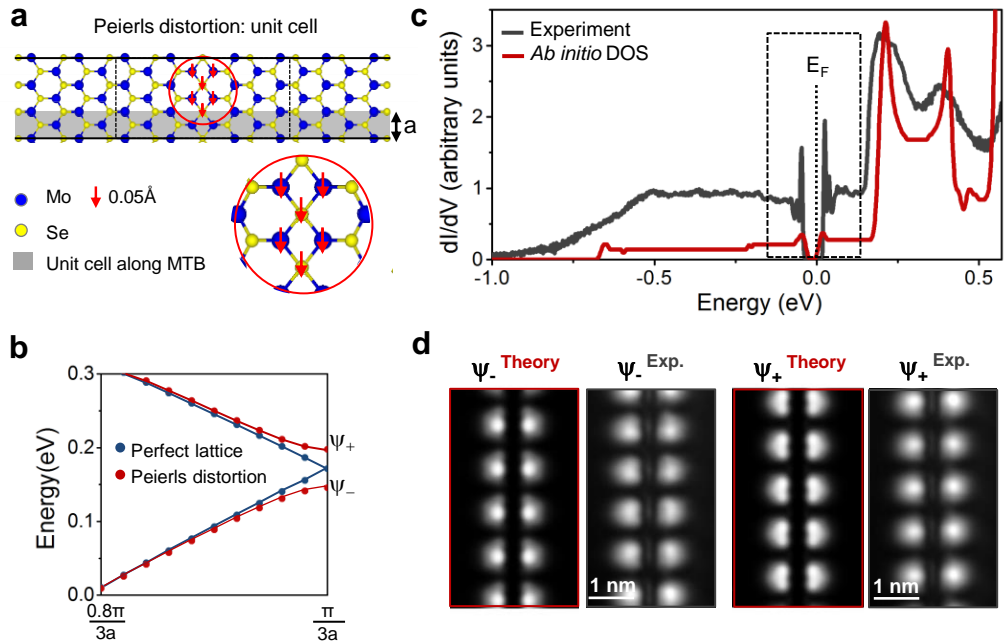

**Figure 4: Effect of a Peierls distortion on the electronic band structure of mirror twin boundaries in monolayer MoSe$_2$ from DFT calculations. a**, Atomic structure of the tripled cell along the MTB (solid black lines). The unit cell along the MTB, with length a, is highlighted in grey. Red arrows indicate the distorted positions of the atoms within the MTB. Zoom-in of the circled region is shown in the right-bottom of the figure. **b**, Close-up view around the edge of the folded Brillouin zone (π/3a) of the calculated electronic band structure of a MTB described in a. It shows a gap opening when a Peierls distortion of 0.05 Å is introduced (red), while no gap exists for the perfect system (blue). The calculated Fermi level is set to 0 eV. Note that the gap opens at 0.175 eV above E$_F$. **c**, Projected density of states (PDOS) showing the gap opening when a Peierls distortion is introduced (red), along with the experimental STM dI/dV spectrum (grey). The calculated PDOS has been shifted by -0.175eV in order to center the gap at the Fermi level (see text). **d**, Local DOS maps of the occupied (Ψ$_-$) and unoccupied (Ψ$_+$) states of the distorted MTB, showing out-of-phase modulations similar to experimental observations in the conductance maps.

**Methods:**

The experiments were carried out on high-quality single layers of MoSe$_2$ grown by molecular beam epitaxy on epitaxial bilayers of graphene (BLG) on 6H-SiC(0001). The structural quality and the coverage of the submonolayer MoSe$_2$ samples were characterized by in-situ reflection high-energy electron diffraction (RHEED), low energy electron diffraction (LEED) and core-level photoemission spectroscopy (PES) at the HERS endstation of beamline 10.0.1, Advance Light Source, Lawrence Berkeley National Laboratory[14].

STM/nc-AFM imaging and STS measurements were performed at T= 4.5K in a commercial Createc ultrahigh vacuum (UHV) system equipped with an STM/qPlus sensor. STS differential conductance (dI/dV) point spectra and spatial maps were measured in constant-height mode using standard lock-in techniques (f = 775 Hz, $V_{r.m.s}$ = 2.1 mV, T = 4.5 K). dI/dV spectra on Au(111) substrate was used as an STS reference for tip calibration. The averaged experimental bandgap (95±10 mV) and satellite peak energy were obtained by statistical analysis of over 24 different MTB on several samples and using different tip apexes. Bandgap widths were measured from peak to peak. Nc-AFM images were recorded by measuring the frequency shift of the qPlus resonator (sensor frequency $f_0$ ≈ 30 kHz, Q ≈ 25000) in constant height mode with an oscillation amplitude of 180 pm. Nc-AFM images were measured at a sample bias $V_s$ = - 50 mV, using a tip functionalized with a single CO molecule[40]. STM/STS data was analyzed and rendered using WSxM software[43].

We performed DFT calculations using the PBE functional[42], with the Vienna Ab-initio Simulation Package (VASP)[44] and projector augmented-wave pseudopotentials. We imposed periodic boundary conditions along the MTB direction, and added 10 Å of vacuum on each side of the structure along the other two directions. The size of the simulation box was 3.33 Å along MTB direction, equivalent to the PBE lattice constant for one-layer $MoSe_2$. The atomic positions were optimized until the Hellmann-Feynman force on each atom was less than 0.04 eV/Å. A 1×18×1 k-mesh was used in the relaxation of the unit cell structure, with 18 k-points sampled along the MTB direction. All electronic structure calculations were performed using optimized geometries and were converged with respect to the length of the structure perpendicular to the MTB direction. The CDW electronic structure was calculated using a 1×18×1 k-mesh for the unit cell, and a 1×6×1 k-mesh for the supercell containing three unit cells. Spin-unpolarized calculations were performed; it has been shown that this particular type of MTB structure[38] are not spin polarized. Spin-orbit interactions do not qualitatively change the band structure. The band structure was calculated using 51 k-points sampled along the MTB direction.

**Acknowledgments:**

We acknowledge Prokop Hapala for assistance with the nc-AFM image simulations. We thank our colleagues at the Molecular Foundry for stimulating discussion and assistance. Work at the Molecular Foundry was supported by the Office of Science, Office of Basic Energy Sciences, of the U.S. Department of Energy under Contract No. DE-AC02-05CH11231 (user proposal #3282) (STM imaging, STM spectroscopy, theoretical simulations, and analysis). A.W-B. and S.W. were supported by the U.S. Department of Energy, Office of Science, Basic Energy Sciences, Scientific User Facilities Division (NSRCs) Early Career Award. S.B. acknowledges fellowship support by the European Union under FP7-PEOPLE-2012-IOF-327581. ALS and SIMES were supported by Office of Basic Energy Science, US DOE, under contract numbers DE-AC02-05CH11231 and DE-AC02-76SF00515, respectively. H. R. acknowledges support from Max Planck Korea/POSTECH Research Initiative of the NRF under Project No. NRF-2011-0031558. M.S. was supported by the Division of Materials Science and Engineering through the Chemical and Mechanical Properties of Surfaces and Interfaces program. Portions of the computational work were done with NERSC resources. M.F.C. acknowledges support from National Science Foundation grant EFMA-1542741 (sample surface preparation development).

1. Lahiri, J. *et al.* An extended defect in graphene as a metallic wire. *Nat. Nanotechnol.* **5,** 326–329 (2010).

2. Yazyev, O. V & Louie, S. G. Electronic transport in polycrystalline graphene. *Nat. Mater.* **9,** 806–809 (2010).

3. Ugeda, M. M., Brihuega, I., Guinea, F. & Gómez-Rodríguez, J. M. Missing atom as a source of carbon magnetism. *Phys. Rev. Lett.* **104,** 096804 (2010).

4. Tsen, A. W. *et al.* Tailoring electrical transport across grain boundaries in polycrystalline graphene. *Science* **336,** 1143–1146 (2012).

5. López-Polín, G. *et al.* Increasing the elastic modulus of graphene by controlled defect creation. *Nat. Phys.* **11,** 26–31 (2015).

6. Mak, K. F., Lee, C., Hone, J., Shan, J. & Heinz, T. F. Atomically thin $MoS_2$: a new direct-gap semiconductor. *Phys. Rev. Lett.* **105,** 136805 (2010).

7. Splendiani, A. *et al.* Emerging photoluminescence in monolayer $MoS_2$. *Nano Lett.* **10,** 1271–1275 (2010).

8. Zhang, Y. *et al.* Direct observation of the transition from indirect to direct bandgap in atomically thin epitaxial $MoSe_2$. *Nat. Nanotechnol.* **9,** 111–115 (2014).

9. Baugher, B. W. H., Churchill, H. O. H., Yang, Y. & Jarillo-Herrero, P. Intrinsic electronic transport properties of high-quality monolayer and bilayer MoS$_2$. *Nano Lett.* **13,** 4212–4216 (2013).

10. Kang, K. *et al.* High-mobility three-atom-thick semiconducting films with wafer-scale homogeneity. *Nature* **520,** 656–660 (2015).

11. Qiu, D. Y., da Jornada, F. H. & Louie, S. G. Optical spectrum of MoS$_2$ : many-body effects and diversity of exciton states. *Phys. Rev. Lett.* **111,** 216805 (2013).

12. Bernardi, M., Palummo, M. & Grossman, J. C. Extraordinary sunlight absorption and one nanometer thick photovoltaics using two-dimensional monolayer materials. *Nano Lett.* **13,** 3664–3670 (2013).

13. Britnell, L. *et al.* Strong light-matter interactions in heterostructures of atomically thin films. *Science* **340,** 1311–1314 (2013).

14. Ugeda, M. M. *et al.* Giant bandgap renormalization and excitonic effects in a monolayer transition metal dichalcogenide semiconductor. *Nat. Mater.* **13,** 1091–1095 (2014).

15. Geim, A. K. & Grigorieva, I. V. Van der Waals heterostructures. *Nature* **499,** 419–425 (2013).

16. Lee, C.-H. *et al.* Atomically thin p–n junctions with van der Waals heterointerfaces. *Nat. Nanotechnol.* **9,** 676–681 (2014).

17. Qiu, H. *et al.* Hopping transport through defect-induced localized states in molybdenum disulphide. *Nat. Commun.* **4,** 2642 (2013).

18. Cai, L. *et al.* Vacancy-induced ferromagnetism of MoS$_2$ nanosheets. *J. Am. Chem. Soc.* **137,** 2622–2627 (2015).

19. Van der Zande, A. M. *et al.* Grains and grain boundaries in highly crystalline monolayer molybdenum disulphide. *Nat. Mater.* **12,** 554–561 (2013).

20. Bao, W. *et al.* Visualizing nanoscale excitonic relaxation properties of disordered edges and grain boundaries in monolayer molybdenum disulfide. *Nat. Commun.* **6,** 7993 (2015).

21. Zhang, Z., Zou, X., Crespi, V. H. & Yakobson, B. I. Intrinsic magnetism of grain boundaries in two-dimensional metal dichalcogenides. *ACS Nano* **7,** 10475–10481 (2013).

22. Lauritsen, J. V *et al.* Hydrodesulfurization reaction pathways on MoS$_2$ nanoclusters revealed by scanning tunneling microscopy. *J. Catal.* **224,** 94–106 (2004).

23. Jaramillo, T. F. *et al.* Identification of active edge sites for electrochemical H$_2$ evolution from MoS$_2$ nanocatalysts. *Science* **317,** 100–102 (2007).

24. Grüner, G. *Density waves in solids*. (Perseus, Cambridge, MA, 2000).

25. Peierls, R. E. *Quantum Theory of Solids*. (Oxford University Press, Oxford, Clarendon, 1955).

26. Peng, J.-P. *et al.* Molecular beam epitaxy growth and scanning tunneling microscopy study of TiSe$_2$ ultrathin films. *Phys. Rev. B* **91,** 121113 (2015).

27. Ugeda, M. M. *et al.* Characterization of collective ground states in single-layer NbSe$_2$. *Nat. Phys.* (2015). doi:10.1038/nphys3527

28. Heeger, A. J., Schrieffer, J. R. & Su, W.-P. Solitons in conducting polymers. *Rev. Mod. Phys.* **60,** 781–850 (1988).

29. Yeom, H. W. *et al.* Instability and charge density wave of metallic quantum chains on a silicon surface. *Phys. Rev. Lett.* **82,** 4898–4901 (1999).

30. Jérome, D. Organic conductors: From charge density wave TTF-TCNQ to superconducting (TMTSF)$_2$PF$_6$. *Chem. Rev.* **104,** 5565–5591 (2004).

31. Shin, J. S., Ryang, K.-D. & Yeom, H. W. Finite-length charge-density waves on terminated atomic wires. *Phys. Rev. B* **85,** 073401 (2012).

32. Cheon, S., Kim, T.-H., Lee, S.-H. & Yeom, H. W. Chiral solitons in a coupled double Peierls chain. *Sicence* **350,** 182–185 (2015).

33. Lehtinen, O. *et al.* Atomic scale microstructure and properties of Se-deficient two-dimensional MoSe$_2$. *ACS Nano* **9,** 3274–3283 (2015).

34. Zou, X., Liu, Y. & Yakobson, B. I. Predicting dislocations and grain boundaries in two-dimensional metal-disulfides from the first principles. *Nano Lett.* **13,** 253–258 (2013).

35. Zhou, W. *et al.* Intrinsic structural defects in monolayer molybdenum disulfide. *Nano Lett.* **13,** 2615–2622 (2013).

36. Liu, H. *et al.* Dense network of one-dimensional midgap metallic modes in monolayer MoSe$_2$ and their spatial undulations. *Phys. Rev. Lett.* **113,** 066105 (2014).

37. Gibertini, M. & Marzari, N. Emergence of one-dimensional wires of free carriers in transition-metal-dichalcogenide nanostructures. *Nano Lett.* **15,** 6229–6238 (2015).

38. Le, D. & Rahman, T. S. Joined edges in MoS$_2$: metallic and half-metallic wires. *J. Phys. Condens. Matter* **25,** 312201 (2013).

39. Murata, H. & Koma, A. Modulated STM images of ultrathin MoSe$_2$ films grown on MoS2(0001) studied by STM/STS. *Phys. Rev. B* **59,** 10327–10334 (1999).

40. Gross, L., Mohn, F., Moll, N., Liljeroth, P. & Meyer, G. The chemical structure of a molecule resolved by atomic force microscopy. *Science* **325,** 1110–1114 (2009).

41. Hapala, P. *et al.* Mechanism of high-resolution STM/AFM imaging with functionalized tips. *Phys. Rev. B* **90,** 085421 (2014).

42. Perdew, J., Burke, K. & Ernzerhof, M. Generalized Gradient Approximation Made Simple. *Phys. Rev. Lett.* **77,** 3865–3868 (1996).

43. Horcas, I. *et al.* WSXM: A software for scanning probe microscopy and a tool for nanotechnology. *Rev. Sci. Instrum.* **78,** 013705 (2007).

44. Kresse, G. & Furthmüller, J. Efficient iterative schemes for ab initio total-energy calculations using a plane-wave basis set. *Phys. Rev. B* **54,** 11169–11186 (1996).

# Supplementary Information

# Observation of charge density wave order in 1D mirror twin boundaries of single-layer MoSe$_2$

1. CDW modulation periodicity: Independence of Moiré pattern periodicities
2. Electronic band structure of a unit cell along the MTB system: formation of a metallic band
3. Electronic band structure of doubled and tripled unit cell along the MTB systems: effect of Peierls distortion

**1. CDW modulation periodicity: Independence of Moiré pattern periodicities**

High-resolution STM images acquired on monolayer MoSe$_2$ grown on bilayer graphene (BLG) show a periodic superstructure (moiré pattern) on top of the atomic lattice periodicity. This moiré pattern is due to the different lattice parameters of MoSe$_2$ and graphene unit cells, as well as the relative stacking orientation between them (angle θ). The most commonly observed moiré pattern for single-layer MoSe$_2$ grown on BLG by molecular beam epitaxy arises from the overlay of the MoSe$_2$ lattice aligned (θ=0°) respect to the BLG lattice. A 4:3

graphene-MoSe$_2$ coincidence gives a quasi-commensurate 9.87Å x 9.87 Å moiré pattern (Fig. S1a). However, the atomic registry of the MoSe$_2$ lattice frequently shows a rotational angle with respect to the graphene lattice and additional moiré pattern periodicities arise (Fig. S1b). As seen in Fig. S1, the modulation of the intensity along the mirror twin boundary (MTB) remains constant (~1 nm) independent of the moiré periodicity. Therefore, we can exclude the superlattice potential induced by the moiré pattern as the origin of the modulation in the intensity along the MTB[1].

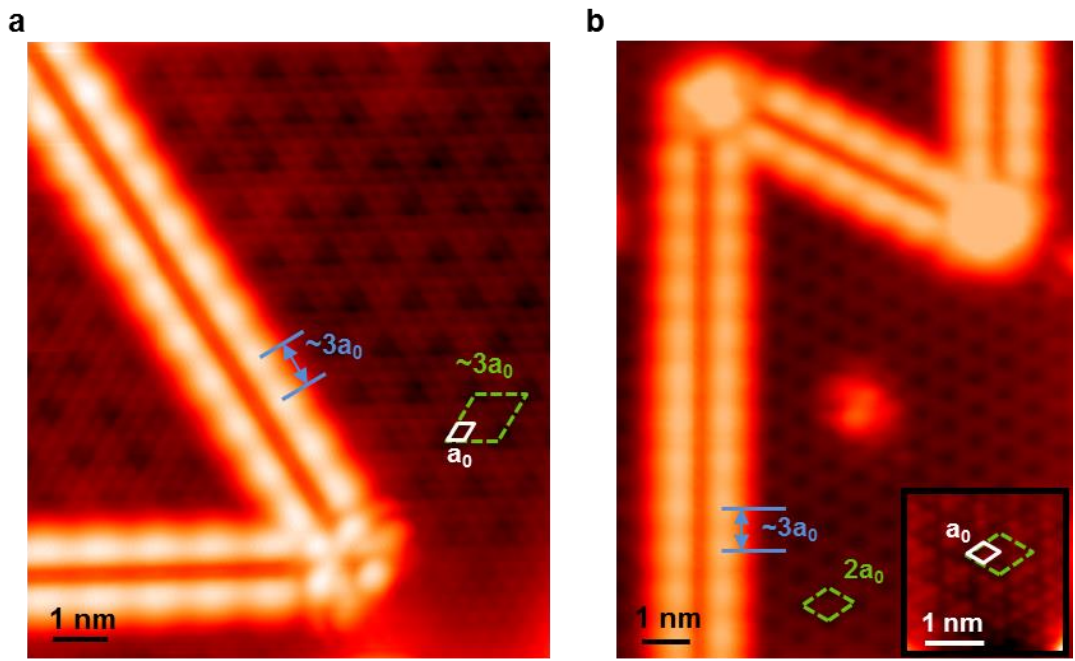

**Figure S1. Moiré pattern periodicities on single layer MoSe$_2$ on bilayer graphene.** High resolution STM images of single layer MoSe$_2$ on BLG presenting moiré patterns with periodicities of **a,** 3$a_0$, and **b,** 2$a_0$, where $a_0$ is the MoSe$_2$ lattice periodicity (Inset: atomic resolution STM image of the same surface region as the main image). The periodicity of the intensity modulation along the MTBs in both images is ~1 nm – approximately three times the atomic lattice periodicity. Approximate moiré unit cells are outlined in green and MoSe$_2$ lattice unit cell in white. ($V_s$ = 1.5 V, $I_t$ = 10 pA, T = 4.5 K)

2. **Electronic band structure of a unit cell along the MTB system: formation of a metallic band**

Non-contact atomic force microscopy was used to develop a structural model for the MTB. We computed the electronic structure of the system using density functional theory (DFT) calculations employing the Perdew-Burke-Ernzerhof (PBE) functional[2]. The unit cell along the MTB is shown in Fig. S2a. We imposed periodic boundary conditions along the MTB, and added 10 Å of vacuum on each side of the structure along the other two directions. The size of the simulation box was 3.33 Å along the MTB, equivalent to the PBE lattice constant for one-layer $MoSe_2$. The atomic positions were optimized until the force on each atom was less than 0.04 eV/Å. All electronic structure calculations were performed using the relaxed structure. We have checked convergence with respect to the length of the structure perpendicular to the MTB direction. There is no difference in the electronic band structure between the computed for the given unit cell structure (see Fig. S2a) and for one whose "bulk" region (defined as the grey box region in Fig. S2a) is four times as wide as the current structure. In the region labeled "bulk" the electronic structure is not perturbed significantly either by the MTB or by the edge of the unit cell structure. The geometry relaxation and electronic structure were calculated using a 1×18×1 k-mesh, with 18 k-points sampled along the MTB. The results shown are from spin-unpolarized calculations, because spin-polarized calculations do not make a significant difference for this particular MTB structure[3]. The band structure calculation was performed non self-consistently with 51 k-points sampled along the MTB direction.

Fig. S2b shows the calculated electronic structure of the unit cell along the MTB system described in Fig. S2a. Grey dots represent the contribution from electronic states spatially localized in the "bulk" (grey-boxed region in Fig. S2a) and red dots represent the contribution from states that are spatially localized in the MTB region, extended 10 Å on each side of the MTB (red-boxed region in Fig. S2a). The MTB gives rise to a metallic band that crosses the Fermi level at slightly less than 1/3 of the way along the Γ-X direction of the Brillouin zone, consistent with previous work[3]. In comparison, Fig. S2c shows the electronic structure of the system including of spin-orbing coupling. One can see that the spin-orbit coupling does not change the band structure significantly for this system, although the degeneracy of states around the midpoint of the Brillouin zone is lifted (blue circles).

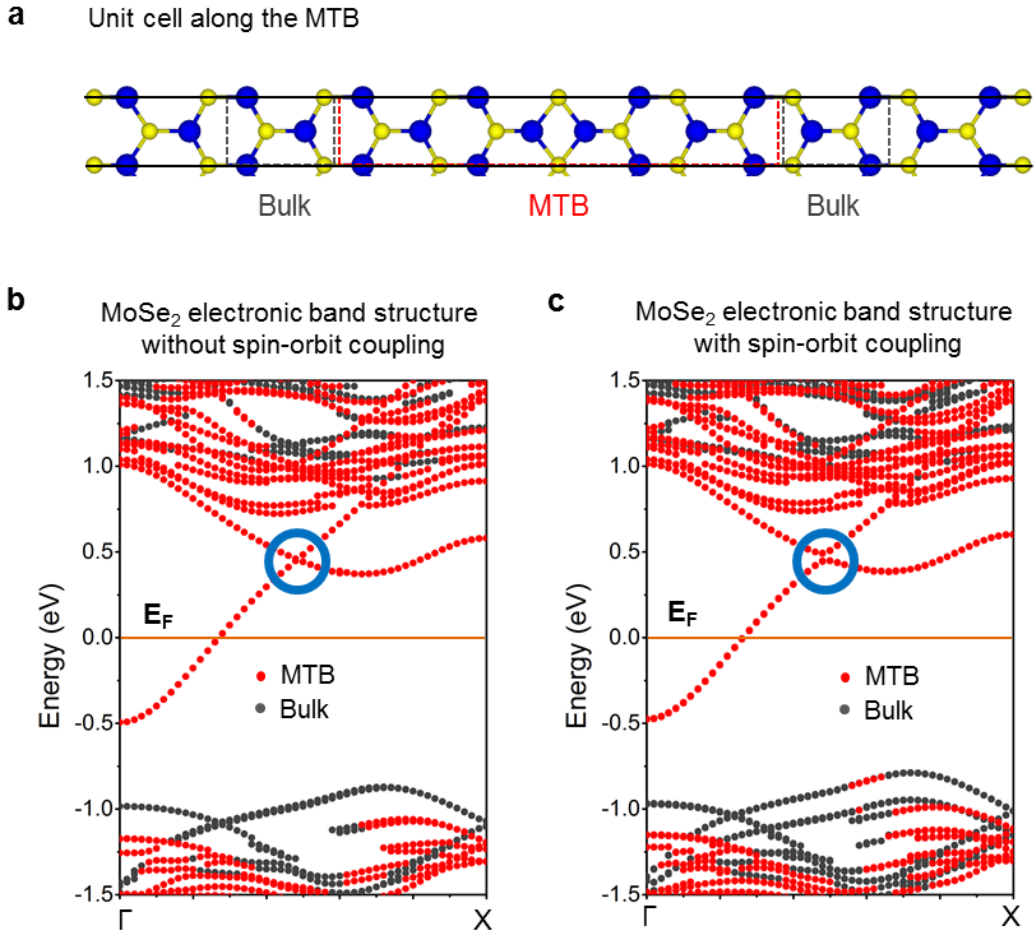

**Figure S2. Electronic band structure of the unit cell along the MTB system from DFT calculations. a,** Atomic structure of the unit cell along the MTB direction (solid black lines). Blue (yellow) spheres represent Mo (Se) atoms. **b,** Electronic band structure of a MTB in single layer $MoSe_2$ with one unit cell along the direction of the MTB, without spin-orbit coupling. Grey (red) dots represent the states spatially localized in the "bulk" (MTB) region boxed in grey (red) in a. The new metallic band crosses the Fermi level (orange line) at slightly less than 1/3 of the $\Gamma$-X distance (Brillouin zone boundary). **c,** Same as b, but with spin-orbit coupling. One can see that the spin-orbit coupling does not change the band structure significantly, although the degeneracy of states around midpoint of the Brillouin zone is lifted (blue circles).

## 3. Electronic band structure of doubled and tripled unit cell along the MTB systems: effect of Peierls distortion

As seen in the previous section there is a band crossing the Fermi level at around 1/3 of the Brillouin zone that suggests the possibility of a charge density wave instability. We therefore studied the effect of a Peierls distortion on the electronic structure of the MTB by a) doubling or tripling the unit cell along the MTB, and b) introducing small distortions with the periodicity of such supercells. Fig. S3a and b show the structural models of the doubled and tripled unit cells. Calculations were performed for a number of different types of distortions with the periodicity of the supercells. The red arrows show one case.

The electronic structure calculations were performed using a 1×9×1 k-mesh for the distorted two-unit-cell system, and a 1×6×1 k-mesh for the distorted three-unit-cell system. The band structure calculations were performed afterwards with 51 k-points sampled along the MTB direction. Fig. S3c shows the electronic structure of a MTB with a doubled unit cell for the perfect lattice (red) and when a Peierls distortion of 0.1 Å is introduced along the line defect (blue). Essentially, Peierls distortions with a periodicity of two times the lattice parameter do not open a gap in the density of states near the Fermi level. On the other hand, as seen in Fig. S3d, if the introduced distortion has a periodicity of three times the lattice parameter, a small gap opens near the Fermi level, at the edge of the Brillouin zone of the three-unit-cell system, consistent with the measured experimental gap. Spin-orbit coupling considerations in the calculations do not yield a qualitatively different band structure near the gap. Both, symmetric and asymmetric distortions in the tripled unit cell lead to qualitatively the same gap opening.

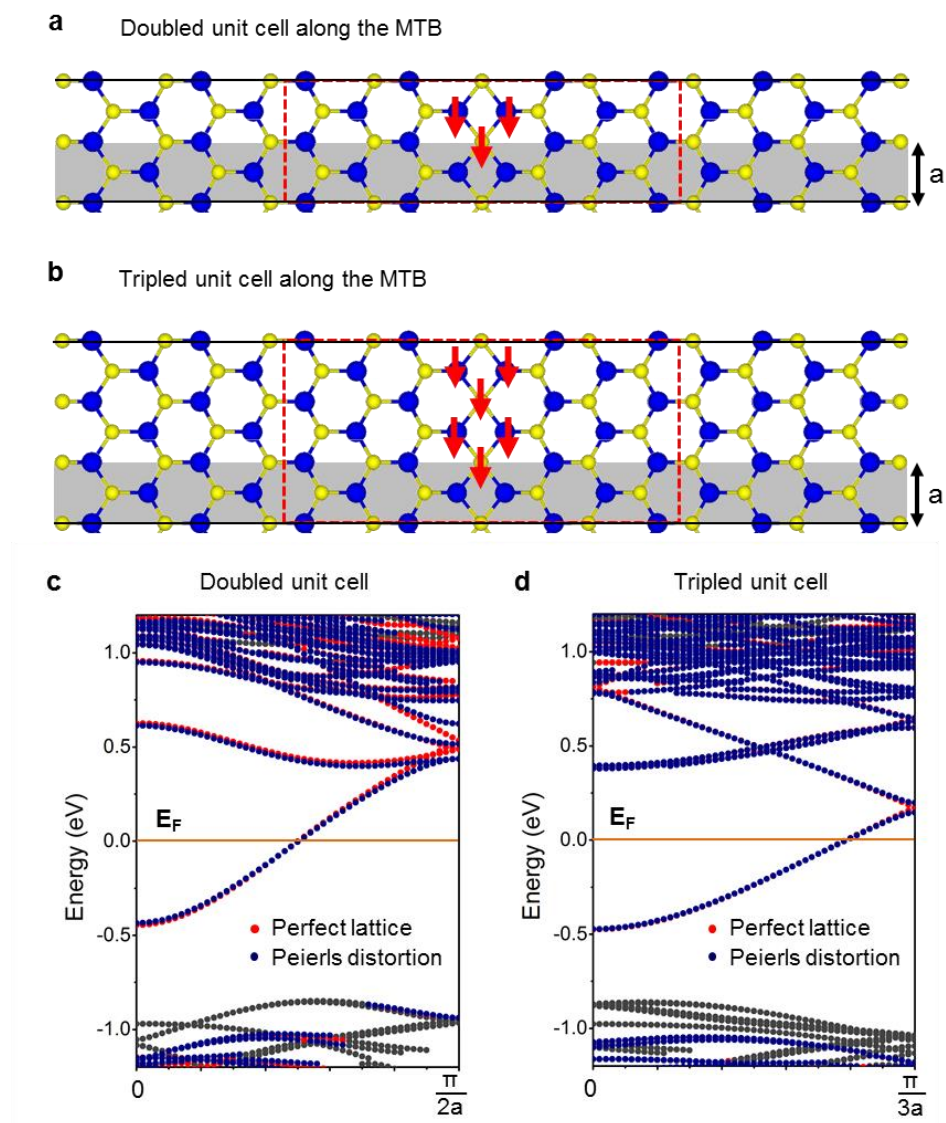

**Figure S3. Effect of a Peierls distortion on the electronic band structure of doubled and tripled unit cell MTB systems from DFT calculations.** Atomic structure of the MTB supercell composed of **a,** two and **b**, three unit cells along the direction of the MTB (solid black lines). Blue (yellow) spheres represent Mo (Se) atoms. **c,** Electronic band structure of a MTB with two unit cells along the MTB. The states localized in the MTB region (red-boxed region in a) are highlighted with red dots for the perfect lattice and blue dots for the system after a Peierls distortion of 0.1 Å. **d,** Electronic band structure of a MTB with three unit cells along the MTB. The states localized in the MTB region are highlighted with red dots for the perfect lattice and blue dots for the system after a Peierls distortion of 0.05 Å. A gap opens at the edge of the Brillouin zone near the Fermi energy (orange line). Grey dots are states in the "bulk".

The supercell distortions do not introduce significant long wavelength modulations for arbitrary states, as shown by the local density of states (LDOS) map (Fig. S4b), calculated for a state away from the gap (red circle in Fig. S4a). For comparison we reproduce the LDOS map from the main text (Fig. S4c) for a state at the edge of the gap (green circle in Fig. S4a).

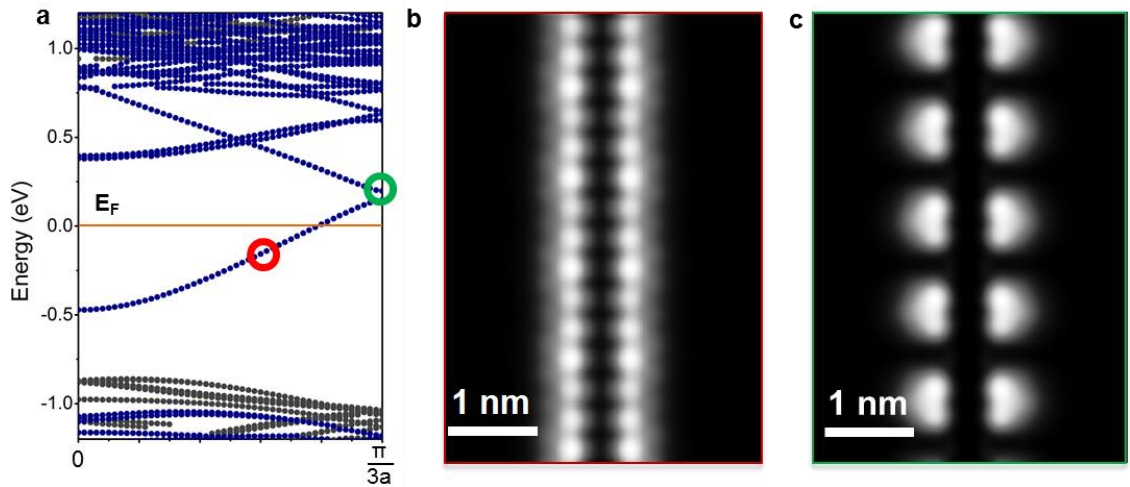

**Fig S4. Calculated LDOS maps for the distorted three unit cell MTB system. a,** The band structure from Fig. **b,** LDOS map of the state indicated in a by a red circle, showing no significant modulation with a periodicity of three times the lattice constant. **c,** LDOS map of the state indicated in a by a green circle, showing modulation with a periodicity of three times the lattice constant.

1. Liu, H. *et al.* Dense network of one-dimensional midgap metallic modes in monolayer $MoSe_2$ and their spatial undulations. *Phys. Rev. Lett.* **113,** 066105 (2014).

2. Perdew, J., Burke, K. & Ernzerhof, M. Generalized Gradient Approximation Made Simple. *Phys. Rev. Lett.* **77,** 3865–3868 (1996).

3. Le, D. & Rahman, T. S. Joined edges in $MoS_2$: metallic and half-metallic wires. *J. Phys. Condens. Matter* **25,** 312201 (2013).